\documentclass[prl,twocolumn,showpacs,superscriptaddress,fleqn,floatfix,footinbib,preprintnumbers,10pt,aps]{revtex4-1}

\pdfoutput=1

\usepackage{graphicx}
\usepackage{amsmath,amssymb}
\usepackage{bm}

\DeclareMathAlphabet{\mathcal}{OMS}{cmsy}{b}{n}

\begin{document}
\preprint{(Accepted for publication in Phys. Rev. Lett.)}

\title{Electronic friction-based vibrational lifetimes of molecular 
adsorbates:\\ Beyond the independent atom approximation}

\author{Simon P. Rittmeyer}
\email[Corresponding author: ]{simon.rittmeyer@tum.de}
\affiliation{Chair for Theoretical Chemistry and Catalysis Research Center, 
Technische Universit\"at M\"unchen, Lichtenbergstr. 4, 85747 Garching, Germany}
\author{J\"org Meyer}
\email{j.meyer@chem.leidenuniv.nl}
\affiliation{Leiden Institute of Chemistry, Gorlaeus Laboratories, Leiden 
University, P.O. Box 9502, 2300 RA Leiden, The Netherlands}
\author{J. I\~naki Juaristi}
\affiliation{Departamento de F\'isica de Materiales, Facultad de Qu\'imicas, 
UPV/EHU, Apartado 1072, 20080 San Sebasti\'an, Spain} 
\affiliation{Centro de F\'isica de Materiales CFM/MPC (CSIC-UPV/EHU), Paseo 
Manuel de Lardizabal 5, 20018 San Sebasti\'an, Spain}
\affiliation{Donostia International Physics Center (DIPC), Paseo Manuel de 
Lardizabal 5, 20018 San Sebasti\'an, Spain}
\author{Karsten Reuter}
\affiliation{Chair for Theoretical Chemistry and Catalysis Research Center, 
Technische Universit\"at M\"unchen, Lichtenbergstr. 4, 85747 Garching, Germany}

\begin{abstract} 
We assess the accuracy of vibrational damping rates of diatomic adsorbates on metal surfaces as calculated within the local-density friction approximation (LDFA). An atoms-in-molecules (AIM) type charge partitioning scheme accounts for intra-molecular contributions and overcomes the systematic underestimation of the non-adiabatic losses obtained within the prevalent independent atom approximation. The quantitative agreement obtained with theoretical and experimental benchmark data suggests the LDFA-AIM as an efficient and reliable approach to account for electronic dissipation in \emph{ab initio} molecular dynamics simulations of surface chemical reactions.
\end{abstract}


\maketitle

A central challenge in energy and catalysis applications is to transfer energy specifically into those degrees of freedom that actually drive a desired surface chemical reaction -- and to keep this energy in these degrees of freedom for a sufficiently long time. In this transfer of energy, losses due to electronic non-adiabaticity can be an important dissipation channel \cite{Wodtke_IntRevPhysChem_2004, Nienhaus_SurfSciRep_2002}. In aiming to assess this channel for systems of technological interest, predictive-quality calculations would be a valuable addition to experimental endeavors. Especially for chemical reactions at frequently employed metal substrates, however, a corresponding methodology has not yet been well established.

To date, most accurate solutions of the full nuclear-electron wave function are restricted to systems of the complexity level of gas-phase ${\text{H}_2}^+$ \cite{Abedi_PhysRevLett_2010}. In the limit of weak non-adiabaticity as pertinent to electron-hole (eh) pair excitations during adsorbate dynamics on metal surfaces, less rigorous approaches rely on mixed quantum-classical dynamics. The imposed computational burden nevertheless still restricts their practical use to simple metals and sub-picosecond time scales \cite{Lindenblatt_PhysRevLett_2006, Grotemeyer_PhysRevLett_2014}, to symmetric adsorbate trajectories \cite{Timmer_PhysRevB_2009, Meyer_NewJPhys_2011}, or to only qualitative accounts of the metal electronic structure \cite{Shenvi_Science_2009, Shenvi_JChemPhys_2009}. Presently, it is thus only the concept of electronic friction \cite{Echenique_SolidStateCommun_1981, Hellsing_PhysScripta_1984, Echenique_PhysRevA_1986, HeadGordon_JChemPhys_1995} and its incorporation into classical molecular dynamics (MD) simulations on the Born-Oppenheimer potential energy surface (PES) $V_{\text{PES}}$ \cite{Trail_JChemPhys_2003, Luntz_JChemPhys_2005, Juaristi_PhysRevLett_2008, Fuechsel_PhysChemChemPhys_2011, MartinGondre_PhysRevLett_2012, Fuechsel_JPhysChemA_2013, BlancoRey_PhysRevLett_2014,Saalfrank_JChemPhys_2014} that promises predictive-quality and material-specific trajectory calculations over an extended period of time.

Particularly the local-density friction approximation (LDFA) \cite{Li_PhysRevLett_1992, Juaristi_PhysRevLett_2008} and for molecular adsorbates an additional independent atom approximation (IAA) \cite{Juaristi_PhysRevLett_2008, Fuechsel_PhysChemChemPhys_2011, MartinGondre_PhysRevLett_2012} provide a further decrease in computational cost. This has allowed for first accounts of electronic non-adiabaticity in large-scale MD simulations based on a first-principles and high-dimensional description of the underlying PES -- either interpolated \cite{Juaristi_PhysRevLett_2008, Fuechsel_PhysChemChemPhys_2011, MartinGondre_PhysRevLett_2012, Fuechsel_JPhysChemA_2013} or even evaluated on-the-fly within \emph{ab initio} MD simulations \cite{BlancoRey_PhysRevLett_2014,Saalfrank_JChemPhys_2014}. However, due to the drastic simplifications introduced with the IAA, the validity of the LDFA formalism for molecular adsorbates \emph{per se} has been controversially discussed \cite{Juaristi_PhysRevLett_2008, Luntz_PhysRevLett_2009, Juaristi_PhysRevLett_2009}. By construction the IAA does not resolve the electronic structure of the interacting molecule-surface system and in particular the location of the molecular frontier orbitals in the surface band structure. It can thus, for instance, not reproduce the enhancement of friction coefficients close to the transition state of a molecular dissociation event on metal surfaces \cite{Luntz_PhysRevLett_2009}. On the other hand, the necessity of an accurate description of such regions of enhanced friction for the overall non-adiabatic energy dissipation has been questioned, as the typically low velocities in these regions effectively suppress the contribution of the friction term within the dynamics \cite{Juaristi_PhysRevLett_2008}.  

Despite the success in recent applications, it thus remains elusive to which extent the limitations of the prevalent IAA carry over to actual observables. In this situation, the vibrational lifetimes of high-frequency adsorbate modes can provide a sensitive measure, as they are largely governed by energy dissipation in the electronic non-adiabatic channel \cite{Tully_JVacSciTechnolA_1993,Arnolds_ProgSurfSci_2011, Saalfrank_ChemRev_2006}. Accurate experimental reference data then allows for a substantiated assessment of the quality of the non-adiabatic description. In this study we perform such an assessment, primarily focusing on the internal stretch mode of two systems which have been studied most extensively and conclusively by experiments: CO adsorbed on Cu(100) and Pt(111) \cite{Arnolds_ProgSurfSci_2011}. Despite the largely different surface frontier orbital locations and concomitant hybridizations at the transition and noble metal surface, we find the LDFA-IAA to already exhibit a good qualitative performance with respect to experimental \cite{Beckerle_JChemPhys_1991, Morin_JChemPhys_1992} and theoretical \cite{Forsblom_JChemPhys_2007, Krishna_JChemPhys_2006} benchmark data. Rather than an explicit account of the surface band structure, our analysis suggests missing intra-molecular contributions as reason for the remaining differences. Approximately incorporating such contributions through a numerically efficient atoms-in-molecules charge partitioning indeed yields consistent lifetimes for a range of diatomic adsorbate systems.

In friction theory, all non-adiabatic effects due to the excitation of eh pairs in the metal substrate are condensed into a single velocity-dependent dissipative force that augments the classical equations of motion
\begin{equation}
    m_i \frac{\text{d}^2{R_{i\alpha}}}{\text{d}t^2} =
		- \frac{\partial V_{\text{PES}}}{\partial R_{i\alpha}} 
    - \sum_{j=1}^N\sum_{\beta = 1}^3 \eta_{i\alpha j\beta}
    \frac{\text{d}R_{j\beta}}{\text{d}t}
    +\mathcal{F}_{i\alpha}(T)	\,.
    \label{eq:EOM_with_friction}
\end{equation}
Here, small latin and greek subscripts denote atoms and Cartesian degrees of freedom, respectively, and $N$ is the total number of atoms of mass $m_i$ and position $\mathbf{R}_i$ in the system. The fluctuating white noise force $\mathcal{F}_{i\alpha}(T)$ becomes negligible at very low temperatures and vanishes exactly at $0\,\text{K}$. Every element of the friction tensor $\bm{\eta}$ is, in principle, a function of all nuclear coordinates. Within the spirit of weak coupling, the focus is usually on the diagonal contributions describing the electronic friction felt by each atom \cite{HeadGordon_JChemPhys_1995, Saalfrank_ChemRev_2006}.

These atomic friction coefficients $\eta_{i\alpha j\beta} = \eta_{i\alpha} \delta_{ij}\delta_{\alpha\beta}$ can e.g. be calculated within the quasi-static regime building on time-dependent density-functional theory (DFT) as suggested by Persson and Hellsing \cite{Hellsing_PhysScripta_1984, Persson_PhysRevLett_1982}. While insightful and generally in good agreement with experiments, the accurate numerical evaluation of this approach is challenging in practice and has hitherto been limited to low-dimensional potentials describing the adsorbate-metal interaction \cite{Lorente_FaradayDiscuss_2000, Trail_JChemPhys_2003, Luntz_JChemPhys_2005, Krishna_JChemPhys_2006, Luntz_JChemPhys_2006}. This shifts interest to more effective schemes and there in particular to the local-density friction approximation. The LDFA introduces isotropic scalar atomic friction coefficients $\eta_{i}^{\text{LDFA}}$, such that $\eta_{i\alpha j\beta}^{\text{LDFA}} = \eta_{i}^{\text{LDFA}} \delta_{ij}\delta_{\alpha\beta}$. These coefficients can be calculated very efficiently from the scattering properties of an atomic impurity embedded in a free electron gas (FEG) \cite{Echenique_SolidStateCommun_1981, Puska_PhysRevB_1983, Echenique_PhysRevA_1986}. In order to relate this model to the actual motion of adsorbates, the embedding density of the FEG $\rho_{\text{emb}}$ is then chosen as the electron density value of the clean metal surface $\rho_{\text{surf}}$ at the position of the adsorbate atoms. In its application to molecular adsorbates this implies an independent atom approximation -- which has been regarded and discussed as being inherently included in the LDFA formalism in this context \cite{Juaristi_PhysRevLett_2008, Luntz_PhysRevLett_2009, Juaristi_PhysRevLett_2009}. As a result of the IAA, the employed atomic friction coefficients are insensitive to the molecular nature of the adsorbate, yet can be evaluated very efficiently. The only input variable is the clean surface electronic density, which, assuming a frozen surface, has to be calculated only once in advance.

We obtain the two-dimensional PES $V_{\text{PES}}(d_{\text{CO}}, Z_{\text{COM}})$ of an adsorbed CO molecule as a function of its bond length $d_{\text{CO}}$ and center-of-mass (COM) height $Z_{\text{COM}}$ above the frozen surface through DFT calculations. Each PES is supported by 442 data points that are calculated with the CASTEP plane-wave pseudopotential code \cite{Clark_ZKristallogr_2005} and subsequently interpolated using bivariate cubic splines. Electronic exchange and correlation (xc) is treated on the GGA level in terms of the PBE functional \cite{Perdew_PhysRevLett_1996, *Perdew_PhysRevLett_1997}. The metal surfaces are modeled by five layer slabs with a separating $20\,\text{\AA}$ vacuum distance. We consider top-site adsorbed CO molecules on one side of the slab within a $c(2\times2)$ and $(\sqrt{3}\times\sqrt{3})\text{R}30^\circ$ surface unit-cell on Cu(100) and Pt(111), respectively. In both cases, the molecular axis is perpendicular to the surface, with the C atom coordinated to the metal atom. At the employed computational settings ($600\,\text{eV}$ cut-off energy, ultrasoft pseudopotentials \cite{Vanderbilt_PhysRevB_1990}, $(10\times10\times1)$ and $(11\times11\times1)$ Monkhorst-Pack $\mathbf{k}$-point grids \cite{Monkhorst_PhysRevB_1976} for Cu(100) and Pt(111), respectively) the PES data points are converged to $<5\,\text{meV}$.  Our investigation is not affected by the well-known CO adsorption puzzle and the concomitant wrong absolute depth of the adsorption well \cite{Feibelman_JPhysChem_2001}. This is confirmed by essentially identical lifetimes we obtain when using PESs generated with a van der Waals-xc-functional \cite{Dion_PhysRevLett_2004} that leads to a stabilization of the top site \cite{Lazic_PhysRevB_2010}. 

Vibrational lifetimes $\tau$ are extracted from classical MD simulations on the interpolated PESs by numerically solving Eq.~(\ref{eq:EOM_with_friction}). Within the LDFA the atomic friction coefficients $\eta^\text{LDFA}_i$ are calculated from the scattering phase shifts of the Kohn-Sham orbitals at the Fermi momentum $\delta_l^{\text{F}} = \delta_l(k_\text{F})$ for an atomic impurity embedded in
a FEG of density $\rho_{\text{emb},i}$ \cite{Echenique_SolidStateCommun_1981, Puska_PhysRevB_1983, Echenique_PhysRevA_1986}
\begin{equation}
    \eta_i^{\text{LDFA}}(\rho_{\text{emb},i})
	 = \frac{4\pi\rho_{\text{emb},i}}{k_\text{F}}\sum_{l=0}^\infty (l+1)\sin^2
	 \left[\delta_l^{\text{F}} - \delta_{l+1}^{\text{F}}\right]
	 \; ,
	 \label{eqn:eta}
\end{equation}
where $\rho_{\text{emb},i}^{\text{IAA}} = \rho_\text{surf}(\mathbf{R}_i)$ within the IAA as described before. Assuming a constant energy dissipation rate and thus an exponential decay of the vibrational energy $E_\text{vib}$, the lifetime $\tau$ can be extracted from the simulations by a logarithmic fit of $E_\text{vib}$ versus time $t$. To initialize our simulations, we assign the adsorbate stretch-mode a projected kinetic energy of $\hbar\omega$, where $\omega$ is the normal mode frequency. Higher initial kinetic energies up to $5\hbar\omega$ result in minute lifetime changes of less than 0.1\,ps.

\begin{figure}
    \includegraphics{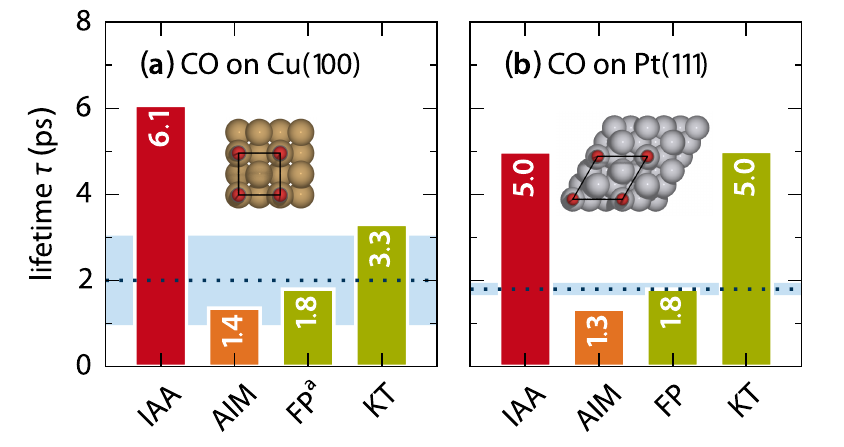}
    \caption{(Color online) Vibrational lifetimes for CO on (a) Cu(100) and (b) Pt(111). Values as obtained within the independent atom approximation (IAA) and the atoms-in-molecules (AIM) approach from Eq.~(\ref{eqn:difference-density}) are contrasted to corresponding predicted lifetimes published by Forsblom and Persson (FP) \cite{Forsblom_JChemPhys_2007} and Krishna and Tully (KT) \cite{Krishna_JChemPhys_2006}. For comparison, experimental values as obtained from pump-probe spectroscopy by Morin \emph{et al.} for CO on Cu(100) \cite{Morin_JChemPhys_1992} and Beckerle \emph{et al.} for CO on Pt(111) \cite{Beckerle_JChemPhys_1991} are shown as a dotted line and a blue stripe further indicating the reported experimental uncertainty. $^\text{a}$CO on Cu(111)}
    \label{fig:lifetimes}
\end{figure}

Figure~\ref{fig:lifetimes} shows the vibrational lifetimes that result from our simulations. We compare them to experimental values obtained from pump-probe spectroscopy \cite{Morin_JChemPhys_1992, Beckerle_JChemPhys_1991} and theoretical values as published by Forsblom and Persson (FP) \cite{Forsblom_JChemPhys_2007}, and Krishna and Tully (KT) \cite{Krishna_JChemPhys_2006}. The latter two are both based on the orbital-dependent Persson/Hellsing expression mentioned above \cite{Persson_PhysRevLett_1982, Hellsing_PhysScripta_1984}, albeit obtained from different derivations and relying on slightly different numerical treatment. For both systems the LDFA-IAA lifetimes agree fairly well with the theoretical values from FP and KT, as well as with experiment. With all numbers falling within one order of magnitude, the current data thus does not further support the harsh criticism the LDFA-IAA was faced with before \cite{Luntz_PhysRevLett_2009}. Generally, they are instead consistent with the good LDFA-IAA performance reported in earlier studies on non-adiabatic energy losses of various ions scattered off metal surfaces \cite{Winter_PhysRevB_2003} and on the vibrational damping of atoms on metal surfaces \cite{Persson_PhysRevLett_1982, Hellsing_PhysScripta_1984, Tremblay_PhysRevB_2010}. Conspicuously, however, in these studies on adsorbate atoms the agreement was even more quantitative and lacked the systematic underestimation (overestimation) of LDFA-IAA non-adiabatic energy losses (lifetimes) apparent in Fig.~\ref{fig:lifetimes}.

\begin{figure}
    \includegraphics{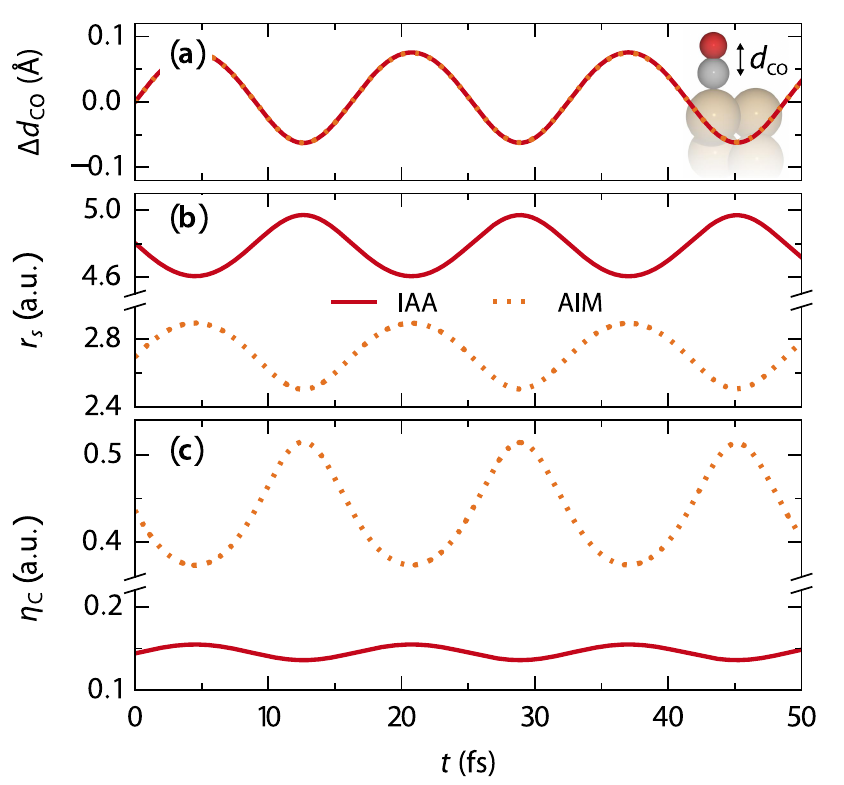}
    \caption{(Color online) (a) Change of the C--O bond length $\Delta d_\text{CO}$ on Cu(100) over the first few periods of the stretch mode. (b) Embedding density $\rho_{\text{emb},\text{C}}$ of the C atom represented by the Wigner-Seitz radius $r_{s} = (3/(4\pi\rho_\text{emb,C}))^{1/3}$ in atomic units (a.u.). (c) Corresponding carbon friction coefficient. Solid red lines refer to the IAA, dotted orange lines to an AIM embedding density according to Eq.~(\ref{eqn:difference-density}).}
    \label{fig:density_and_friction}
\end{figure}

Rather than from a generally insufficient account of the electronic structure of the interacting adsorbate-surface system, this suggests that LDFA-IAA deficiencies arise particularly in the treatment of molecular adsorbates as isolated adatoms.
In this respect -- within the underlying atomic embedding model --
systematic shortcomings lie in the complete neglect of both 
adsorbate-substrate as well as intra-molecular contributions to  $\rho_{\text{emb}}$. Considering only the density of the clean surface, $\rho_{\text{emb}}$ is systematically underestimated; consistent with the underestimated energy losses observed in Fig.~\ref{fig:lifetimes}. Even more, by construction a thus defined $\rho_{\text{emb}}$ is also blind to dynamical changes of intra-molecular bond distances and characters. Figure~\ref{fig:density_and_friction} illustrates this for the vibrational motion of CO at Cu(100). The LDFA-IAA friction coefficient of the carbon atom decreases with smaller C--O bond lengths $d_\text{CO}$ and thus larger distances $Z_{\text{C}}$ of the C atom from the metal surface. This is due to the fact that the IAA only accounts for changes of $\rho_{\text{surf}}$ along the vibrational coordinate. The LDFA-IAA thus effectively decomposes the molecular vibration into two independent adatom vibrations. Yet, even without any intra-molecular bonding there will be an increasing overlap of the O and C atomic densities with decreasing $d_\text{CO}$. This intra-molecular contribution to $\rho_{\text{emb}}$ at the position of every constituent atom is thus completely missed in the LDFA-IAA approximation.

A straightforward way to account for such contributions is to perform a charge decomposition through a projection scheme like Hirshfeld's analysis \cite{Hirshfeld_TheoretChimActa_1977}. For any given adsorbate configuration this provides the projected density of any adsorbate atom $\rho_i^{\text{H}}$, and corresponding sharing function $w_i^{\text{H}}$ at any position $\mathbf{R}$ \cite{McNellis_PhysRevB_2009}. Within an atoms-in-molecules (AIM) picture we can then define as embedding density for atom $i$ at position $\mathbf{R}_i$
\begin{eqnarray}
    \rho^{\text{AIM}}_{\text{emb},i} & = &
    \rho_\text{SCF}\left(\mathbf{R}_i\right) -
    \rho_{i}^{\text{H}}\left(\mathbf{R}_i\right)\notag \\
    & = & \left[ 1 - w_i^{\text{H}}(\mathbf{R}_i) \right]
    \rho_\text{SCF}(\mathbf{R}_i) \;.
    \label{eqn:difference-density}
\end{eqnarray}
We thus consider as embedding density the full self-consistently calculated density $\rho_\text{SCF}$ of the entire interacting adsorbate-surface system just without the contribution associated with atom $i$. This naturally contains density contributions from all other atoms in the system (both substrate and adsorbate) and consequently carries an implicit dependence $\rho^{\text{AIM}}_{\text{emb},i}(\{\mathbf{R}_{i \neq j}\})$. Figure~\ref{fig:density_and_friction}(b) contrasts the correspondingly obtained $\rho^{\text{AIM}}_{\text{emb},\text{C}}$ to $\rho^{\text{IAA}}_{\text{emb},\text{C}}$ during the CO vibrational motion.  Intriguingly, the two quantities differ not only quantitatively, but even exhibit a reversed phase behavior. The account of the density contribution of the O atom in $\rho^\text{AIM}_{\text{emb,C}}$ thus outweighs the influence of the clean surface density seen by the IAA: At the largest vertical heights of the carbon atom $Z_{\text{C}}$, where $\rho^{\text{IAA}}_{\text{emb,C}}$ is smallest, also $d_{\text{CO}}$ is smallest. Already in the simple diatomic this leads to such a large intra-molecular density contribution that the overall $\rho^{\text{AIM}}_{\text{emb,C}}$ is largest.

These new features also carry over to the corresponding friction coefficients, and in turn to the calculated lifetimes. As apparent from Fig.~\ref{fig:lifetimes} the LDFA-AIM approach cures the systematic overestimation of lifetimes and yields values that are now \emph{en par} with the FP calculations and experiments. Similar findings have been obtained when applying the LDFA-AIM to two further systems \footnote{Consistent with the theoretical work by FP \cite{Forsblom_JChemPhys_2007}, we consider perpendicular CN adsorption with the C atom coordinated to the metal on a top site within a $(2\times2)$ surface unit cell. The H$_2$ on Ru(0001) system is modeled by analogy with the equilibrium configuration mentioned by F{\"u}chsel \emph{et al.}. \cite{Fuechsel_JPhysChemA_2013}, i.e., using a $(2\times2)$ surface unit cell with two H atoms adsorbed in adjacent fcc hollow sites. For both systems, we use a $(9\times9\times1)$ Monkhorst-Pack $\mathbf{k}$-point grid \cite{Monkhorst_PhysRevB_1976}. All remaining computational details are the same as outlined for CO on Cu(100) and Pt(111), respectively.} for which reference lifetimes have been reported from orbital-dependent theories, CN on Pt(111) \cite{Forsblom_JChemPhys_2007} and H$_2$ on Ru(0001) \cite{Fuechsel_PhysChemChemPhys_2011}. With $\tau^\text{AIM} = 0.9\,\text{ps}$ vs. $\tau^\text{FP} = 2.4\,\text{ps}$ and $\tau^\text{AIM} = 210\,\text{fs}$ vs. $\tau^\text{\cite{Fuechsel_PhysChemChemPhys_2011}} = 215\,\text{fs}$, respectively, LDFA-AIM yields in both cases lifetimes that are fully consistent with the reference numbers, yet at a fraction of the numerical cost of the orbital-dependent theories. 

Adjusting $\rho_{\text{emb}}$ to also take into account influences from atoms other than only the clean metal surface hence seems to provide a simple, but effective correction for a molecular treatment within the LDFA. The idea underlying Eq.~(\ref{eqn:difference-density}) is thereby similar to the subtraction of a free atom density as suggested in the context of vibrational damping of adatoms \cite{Tremblay_PhysRevB_2010}. However, employing Hirshfeld sharing functions \cite{Hirshfeld_TheoretChimActa_1977} offers the advantage that the embedding densities are guaranteed to be within physically well defined boundaries $0 \leq \rho^{\text{AIM}}_{\text{emb},i} \leq \rho_\text{SCF}(\mathbf{R}_i)$. This way, our proposed scheme also preserves the molecular dissociation limit by construction: At large bond distances, the respective friction coefficients smoothly go over into friction coefficients virtually identical to the ones obtained for independent atoms. This is nicely illustrated by essentially identical vibrational lifetimes obtained within the IAA and AIM for the H$_2$ on Ru(0001) system, where the individual atoms are separated by about $2.7\,\text{\AA}$ on the surface. Furthermore, a Hirshfeld analysis typically requires only a minute computational effort compared to achieving self-consistency for electronic energies and forces (or compared to the multiple self-consistent calculations required to obtain derivatives from finite differences in the FP or KT approaches). The LDFA-AIM scheme proposed here can thus be easily carried out at every time step of \emph{ab initio} MD simulations, allowing surface motion to be explicitly taken into account. In this respect, Eq.~(\ref{eqn:difference-density}) defines an embedding density not only for adsorbate but also for bulk atoms. Friction coefficients derived therefrom could thus, in principle, as well be invoked to evaluate energy losses due to electron-phonon coupling in the bulk.

In conclusion, we have shown that the vibrational damping of high-frequency adsorbate modes on metal surfaces can be added to the list of non-adiabatic phenomena that are reasonably well described by means of electronic friction. The complete neglect of intra-molecular effects in the prevalent LDFA-IAA approximation is thereby likely to underestimate the electronic energy dissipation. The here presented atoms-in-molecules alternative instead accounts for them approximately through a charge partitioning scheme. As it thus effectively treats the molecular electrons as part of the metallic substrate, we expect the AIM friction concept to generally rather overestimate non-adiabatic energy losses and to perform best for chemisorbed adsorbates at close distances to the surface. Being a direct descendant of the LDFA, our scheme is, of course, also unlikely to overcome fundamental limitations that come with its heritage. As such it is unlikely to properly capture the strong enhancement of friction coefficients directly at the transition state leading to molecular dissociation. However, as important as it may be, this actual dissociation event only constitutes a fraction of the dynamics that is relevant to chemical surface reactions. Other important aspects like a vibrational pre-excitation or an ensuing hot adatom motion may dominate the overall (non-adiabatic) energy dissipation \cite{BlancoRey_PhysRevLett_2014, Meyer_AngewChemIntEd_2014}, yet take place over much longer time-scales. They can thus only be assessed with a numerically highly efficient method like the LDFA. In this respect, our results further consolidate the trust in LDFA-based results for these important long-term events. 

\begin{acknowledgments}
JM is grateful to Profs. Mats Persson and Peter Saalfrank for stimulating discussions and thanks Prof. Geert-Jan Kroes for his support through ERC-2013 advanced grant Nr. 338580. J.I.J. acknowledges the Basque Departamento de Educaci\'{o}n, Universidades e Investigaci\'{o}n (Grant No. IT-756-13) and the Spanish Ministerio de Econom\'{i}a y Competitividad (Grant No.  FIS2013-48286-C2-8752-P).  
\end{acknowledgments}

\bibliography{literature}

\begin{thebibliography}{49}%
\makeatletter
\providecommand \@ifxundefined [1]{%
 \@ifx{#1\undefined}
}%
\providecommand \@ifnum [1]{%
 \ifnum #1\expandafter \@firstoftwo
 \else \expandafter \@secondoftwo
 \fi
}%
\providecommand \@ifx [1]{%
 \ifx #1\expandafter \@firstoftwo
 \else \expandafter \@secondoftwo
 \fi
}%
\providecommand \natexlab [1]{#1}%
\providecommand \enquote  [1]{``#1''}%
\providecommand \bibnamefont  [1]{#1}%
\providecommand \bibfnamefont [1]{#1}%
\providecommand \citenamefont [1]{#1}%
\providecommand \href@noop [0]{\@secondoftwo}%
\providecommand \href [0]{\begingroup \@sanitize@url \@href}%
\providecommand \@href[1]{\@@startlink{#1}\@@href}%
\providecommand \@@href[1]{\endgroup#1\@@endlink}%
\providecommand \@sanitize@url [0]{\catcode `\\12\catcode `\$12\catcode
  `\&12\catcode `\#12\catcode `\^12\catcode `\_12\catcode `\%12\relax}%
\providecommand \@@startlink[1]{}%
\providecommand \@@endlink[0]{}%
\providecommand \url  [0]{\begingroup\@sanitize@url \@url }%
\providecommand \@url [1]{\endgroup\@href {#1}{\urlprefix }}%
\providecommand \urlprefix  [0]{URL }%
\providecommand \Eprint [0]{\href }%
\providecommand \doibase [0]{http://dx.doi.org/}%
\providecommand \selectlanguage [0]{\@gobble}%
\providecommand \bibinfo  [0]{\@secondoftwo}%
\providecommand \bibfield  [0]{\@secondoftwo}%
\providecommand \translation [1]{[#1]}%
\providecommand \BibitemOpen [0]{}%
\providecommand \bibitemStop [0]{}%
\providecommand \bibitemNoStop [0]{.\EOS\space}%
\providecommand \EOS [0]{\spacefactor3000\relax}%
\providecommand \BibitemShut  [1]{\csname bibitem#1\endcsname}%
\let\auto@bib@innerbib\@empty
\bibitem [{\citenamefont {Wodtke}\ \emph {et~al.}(2004)\citenamefont {Wodtke},
  \citenamefont {Tully},\ and\ \citenamefont
  {Auerbach}}]{Wodtke_IntRevPhysChem_2004}%
  \BibitemOpen
  \bibfield  {author} {\bibinfo {author} {\bibfnamefont {A.~M.}\ \bibnamefont
  {Wodtke}}, \bibinfo {author} {\bibfnamefont {J.~C.}\ \bibnamefont {Tully}}, \
  and\ \bibinfo {author} {\bibfnamefont {D.~J.}\ \bibnamefont {Auerbach}},\
  }\href {\doibase 10.1080/01442350500037521} {\bibfield  {journal} {\bibinfo
  {journal} {Int. Rev. Phys. Chem.}\ }\textbf {\bibinfo {volume} {23}},\
  \bibinfo {pages} {513} (\bibinfo {year} {2004})}\BibitemShut {NoStop}%
\bibitem [{\citenamefont {Nienhaus}(2002)}]{Nienhaus_SurfSciRep_2002}%
  \BibitemOpen
  \bibfield  {author} {\bibinfo {author} {\bibfnamefont {H.}~\bibnamefont
  {Nienhaus}},\ }\href {\doibase 10.1016/S0167-5729(01)00019-X} {\bibfield
  {journal} {\bibinfo  {journal} {Surf. Sci. Rep.}\ }\textbf {\bibinfo {volume}
  {45}},\ \bibinfo {pages} {1 } (\bibinfo {year} {2002})}\BibitemShut {NoStop}%
\bibitem [{\citenamefont {Abedi}\ \emph {et~al.}(2010)\citenamefont {Abedi},
  \citenamefont {Maitra},\ and\ \citenamefont
  {Gross}}]{Abedi_PhysRevLett_2010}%
  \BibitemOpen
  \bibfield  {author} {\bibinfo {author} {\bibfnamefont {A.}~\bibnamefont
  {Abedi}}, \bibinfo {author} {\bibfnamefont {N.~T.}\ \bibnamefont {Maitra}}, \
  and\ \bibinfo {author} {\bibfnamefont {E.~K.~U.}\ \bibnamefont {Gross}},\
  }\href {\doibase 10.1103/PhysRevLett.105.123002} {\bibfield  {journal}
  {\bibinfo  {journal} {Phys. Rev. Lett.}\ }\textbf {\bibinfo {volume} {105}},\
  \bibinfo {pages} {123002} (\bibinfo {year} {2010})}\BibitemShut {NoStop}%
\bibitem [{\citenamefont {Lindenblatt}\ and\ \citenamefont
  {Pehlke}(2006)}]{Lindenblatt_PhysRevLett_2006}%
  \BibitemOpen
  \bibfield  {author} {\bibinfo {author} {\bibfnamefont {M.}~\bibnamefont
  {Lindenblatt}}\ and\ \bibinfo {author} {\bibfnamefont {E.}~\bibnamefont
  {Pehlke}},\ }\href {\doibase 10.1103/PhysRevLett.97.216101} {\bibfield
  {journal} {\bibinfo  {journal} {Phys. Rev. Lett.}\ }\textbf {\bibinfo
  {volume} {97}},\ \bibinfo {pages} {216101} (\bibinfo {year}
  {2006})}\BibitemShut {NoStop}%
\bibitem [{\citenamefont {Grotemeyer}\ and\ \citenamefont
  {Pehlke}(2014)}]{Grotemeyer_PhysRevLett_2014}%
  \BibitemOpen
  \bibfield  {author} {\bibinfo {author} {\bibfnamefont {M.}~\bibnamefont
  {Grotemeyer}}\ and\ \bibinfo {author} {\bibfnamefont {E.}~\bibnamefont
  {Pehlke}},\ }\href {\doibase 10.1103/PhysRevLett.112.043201} {\bibfield
  {journal} {\bibinfo  {journal} {Phys. Rev. Lett.}\ }\textbf {\bibinfo
  {volume} {112}},\ \bibinfo {pages} {043201} (\bibinfo {year}
  {2014})}\BibitemShut {NoStop}%
\bibitem [{\citenamefont {Timmer}\ and\ \citenamefont
  {Kratzer}(2009)}]{Timmer_PhysRevB_2009}%
  \BibitemOpen
  \bibfield  {author} {\bibinfo {author} {\bibfnamefont {M.}~\bibnamefont
  {Timmer}}\ and\ \bibinfo {author} {\bibfnamefont {P.}~\bibnamefont
  {Kratzer}},\ }\href {\doibase 10.1103/PhysRevB.79.165407} {\bibfield
  {journal} {\bibinfo  {journal} {Phys. Rev. B}\ }\textbf {\bibinfo {volume}
  {79}},\ \bibinfo {pages} {165407} (\bibinfo {year} {2009})}\BibitemShut
  {NoStop}%
\bibitem [{\citenamefont {Meyer}\ and\ \citenamefont
  {Reuter}(2011)}]{Meyer_NewJPhys_2011}%
  \BibitemOpen
  \bibfield  {author} {\bibinfo {author} {\bibfnamefont {J.}~\bibnamefont
  {Meyer}}\ and\ \bibinfo {author} {\bibfnamefont {K.}~\bibnamefont {Reuter}},\
  }\href {\doibase 10.1088/1367-2630/13/8/085010} {\bibfield  {journal}
  {\bibinfo  {journal} {New J. Phys.}\ }\textbf {\bibinfo {volume} {13}},\
  \bibinfo {pages} {085010} (\bibinfo {year} {2011})}\BibitemShut {NoStop}%
\bibitem [{\citenamefont {Shenvi}\ \emph
  {et~al.}(2009{\natexlab{a}})\citenamefont {Shenvi}, \citenamefont {Roy},\
  and\ \citenamefont {Tully}}]{Shenvi_Science_2009}%
  \BibitemOpen
  \bibfield  {author} {\bibinfo {author} {\bibfnamefont {N.}~\bibnamefont
  {Shenvi}}, \bibinfo {author} {\bibfnamefont {S.}~\bibnamefont {Roy}}, \ and\
  \bibinfo {author} {\bibfnamefont {J.~C.}\ \bibnamefont {Tully}},\ }\href
  {\doibase 10.1126/science.1179240} {\bibfield  {journal} {\bibinfo  {journal}
  {Science}\ }\textbf {\bibinfo {volume} {326}},\ \bibinfo {pages} {829}
  (\bibinfo {year} {2009}{\natexlab{a}})}\BibitemShut {NoStop}%
\bibitem [{\citenamefont {Shenvi}\ \emph
  {et~al.}(2009{\natexlab{b}})\citenamefont {Shenvi}, \citenamefont {Roy},\
  and\ \citenamefont {Tully}}]{Shenvi_JChemPhys_2009}%
  \BibitemOpen
  \bibfield  {author} {\bibinfo {author} {\bibfnamefont {N.}~\bibnamefont
  {Shenvi}}, \bibinfo {author} {\bibfnamefont {S.}~\bibnamefont {Roy}}, \ and\
  \bibinfo {author} {\bibfnamefont {J.~C.}\ \bibnamefont {Tully}},\ }\href
  {\doibase 10.1063/1.3125436} {\bibfield  {journal} {\bibinfo  {journal} {J.
  Chem. Phys.}\ }\textbf {\bibinfo {volume} {130}},\ \bibinfo {pages} {174107}
  (\bibinfo {year} {2009}{\natexlab{b}})}\BibitemShut {NoStop}%
\bibitem [{\citenamefont {Echenique}\ \emph {et~al.}(1981)\citenamefont
  {Echenique}, \citenamefont {Nieminen},\ and\ \citenamefont
  {Ritchie}}]{Echenique_SolidStateCommun_1981}%
  \BibitemOpen
  \bibfield  {author} {\bibinfo {author} {\bibfnamefont {P.}~\bibnamefont
  {Echenique}}, \bibinfo {author} {\bibfnamefont {R.}~\bibnamefont {Nieminen}},
  \ and\ \bibinfo {author} {\bibfnamefont {R.}~\bibnamefont {Ritchie}},\ }\href
  {\doibase http://dx.doi.org/10.1016/0038-1098(81)91173-X} {\bibfield
  {journal} {\bibinfo  {journal} {Solid State Commun.}\ }\textbf {\bibinfo
  {volume} {37}},\ \bibinfo {pages} {779 } (\bibinfo {year}
  {1981})}\BibitemShut {NoStop}%
\bibitem [{\citenamefont {Hellsing}\ and\ \citenamefont
  {Persson}(1984)}]{Hellsing_PhysScripta_1984}%
  \BibitemOpen
  \bibfield  {author} {\bibinfo {author} {\bibfnamefont {B.}~\bibnamefont
  {Hellsing}}\ and\ \bibinfo {author} {\bibfnamefont {M.}~\bibnamefont
  {Persson}},\ }\href {\doibase 10.1088/0031-8949/29/4/014} {\bibfield
  {journal} {\bibinfo  {journal} {Phys. Scr.}\ }\textbf {\bibinfo {volume}
  {29}},\ \bibinfo {pages} {360} (\bibinfo {year} {1984})}\BibitemShut
  {NoStop}%
\bibitem [{\citenamefont {Echenique}\ \emph {et~al.}(1986)\citenamefont
  {Echenique}, \citenamefont {Nieminen}, \citenamefont {Ashley},\ and\
  \citenamefont {Ritchie}}]{Echenique_PhysRevA_1986}%
  \BibitemOpen
  \bibfield  {author} {\bibinfo {author} {\bibfnamefont {P.~M.}\ \bibnamefont
  {Echenique}}, \bibinfo {author} {\bibfnamefont {R.~M.}\ \bibnamefont
  {Nieminen}}, \bibinfo {author} {\bibfnamefont {J.~C.}\ \bibnamefont
  {Ashley}}, \ and\ \bibinfo {author} {\bibfnamefont {R.~H.}\ \bibnamefont
  {Ritchie}},\ }\href {\doibase 10.1103/PhysRevA.33.897} {\bibfield  {journal}
  {\bibinfo  {journal} {Phys. Rev. A}\ }\textbf {\bibinfo {volume} {33}},\
  \bibinfo {pages} {897} (\bibinfo {year} {1986})}\BibitemShut {NoStop}%
\bibitem [{\citenamefont {Head-Gordon}\ and\ \citenamefont
  {Tully}(1995)}]{HeadGordon_JChemPhys_1995}%
  \BibitemOpen
  \bibfield  {author} {\bibinfo {author} {\bibfnamefont {M.}~\bibnamefont
  {Head-Gordon}}\ and\ \bibinfo {author} {\bibfnamefont {J.~C.}\ \bibnamefont
  {Tully}},\ }\href {\doibase 10.1063/1.469915} {\bibfield  {journal} {\bibinfo
   {journal} {J. Chem. Phys.}\ }\textbf {\bibinfo {volume} {103}},\ \bibinfo
  {pages} {10137} (\bibinfo {year} {1995})}\BibitemShut {NoStop}%
\bibitem [{\citenamefont {Trail}\ \emph {et~al.}(2003)\citenamefont {Trail},
  \citenamefont {Bird}, \citenamefont {Persson},\ and\ \citenamefont
  {Holloway}}]{Trail_JChemPhys_2003}%
  \BibitemOpen
  \bibfield  {author} {\bibinfo {author} {\bibfnamefont {J.~R.}\ \bibnamefont
  {Trail}}, \bibinfo {author} {\bibfnamefont {D.~M.}\ \bibnamefont {Bird}},
  \bibinfo {author} {\bibfnamefont {M.}~\bibnamefont {Persson}}, \ and\
  \bibinfo {author} {\bibfnamefont {S.}~\bibnamefont {Holloway}},\ }\href
  {\doibase 10.1063/1.1593631} {\bibfield  {journal} {\bibinfo  {journal} {J.
  Chem. Phys.}\ }\textbf {\bibinfo {volume} {119}},\ \bibinfo {pages} {4539}
  (\bibinfo {year} {2003})}\BibitemShut {NoStop}%
\bibitem [{\citenamefont {Luntz}\ and\ \citenamefont
  {Persson}(2005)}]{Luntz_JChemPhys_2005}%
  \BibitemOpen
  \bibfield  {author} {\bibinfo {author} {\bibfnamefont {A.~C.}\ \bibnamefont
  {Luntz}}\ and\ \bibinfo {author} {\bibfnamefont {M.}~\bibnamefont
  {Persson}},\ }\href {\doibase 10.1063/1.2000249} {\bibfield  {journal}
  {\bibinfo  {journal} {J. Chem. Phys.}\ }\textbf {\bibinfo {volume} {123}},\
  \bibinfo {eid} {074704} (\bibinfo {year} {2005})}\BibitemShut {NoStop}%
\bibitem [{\citenamefont {Juaristi}\ \emph {et~al.}(2008)\citenamefont
  {Juaristi}, \citenamefont {Alducin}, \citenamefont {{D\'{i}ez Mui\~no}},
  \citenamefont {Busnengo},\ and\ \citenamefont
  {Salin}}]{Juaristi_PhysRevLett_2008}%
  \BibitemOpen
  \bibfield  {author} {\bibinfo {author} {\bibfnamefont {J.~I.}\ \bibnamefont
  {Juaristi}}, \bibinfo {author} {\bibfnamefont {M.}~\bibnamefont {Alducin}},
  \bibinfo {author} {\bibfnamefont {R.}~\bibnamefont {{D\'{i}ez Mui\~no}}},
  \bibinfo {author} {\bibfnamefont {H.~F.}\ \bibnamefont {Busnengo}}, \ and\
  \bibinfo {author} {\bibfnamefont {A.}~\bibnamefont {Salin}},\ }\href
  {\doibase 10.1103/PhysRevLett.100.116102} {\bibfield  {journal} {\bibinfo
  {journal} {Phys. Rev. Lett.}\ }\textbf {\bibinfo {volume} {100}},\ \bibinfo
  {pages} {116102} (\bibinfo {year} {2008})}\BibitemShut {NoStop}%
\bibitem [{\citenamefont {F{\"u}chsel}\ \emph {et~al.}(2011)\citenamefont
  {F{\"u}chsel}, \citenamefont {Klamroth}, \citenamefont {Monturet},\ and\
  \citenamefont {Saalfrank}}]{Fuechsel_PhysChemChemPhys_2011}%
  \BibitemOpen
  \bibfield  {author} {\bibinfo {author} {\bibfnamefont {G.}~\bibnamefont
  {F{\"u}chsel}}, \bibinfo {author} {\bibfnamefont {T.}~\bibnamefont
  {Klamroth}}, \bibinfo {author} {\bibfnamefont {S.}~\bibnamefont {Monturet}},
  \ and\ \bibinfo {author} {\bibfnamefont {P.}~\bibnamefont {Saalfrank}},\
  }\href {\doibase 10.1039/C0CP02086A} {\bibfield  {journal} {\bibinfo
  {journal} {Phys. Chem. Chem. Phys.}\ }\textbf {\bibinfo {volume} {13}},\
  \bibinfo {pages} {8659} (\bibinfo {year} {2011})}\BibitemShut {NoStop}%
\bibitem [{\citenamefont {Martin-Gondre}\ \emph {et~al.}(2012)\citenamefont
  {Martin-Gondre}, \citenamefont {Alducin}, \citenamefont {Bocan},
  \citenamefont {{D{\'\i}ez Mui{\~n}o}},\ and\ \citenamefont
  {Juaristi}}]{MartinGondre_PhysRevLett_2012}%
  \BibitemOpen
  \bibfield  {author} {\bibinfo {author} {\bibfnamefont {L.}~\bibnamefont
  {Martin-Gondre}}, \bibinfo {author} {\bibfnamefont {M.}~\bibnamefont
  {Alducin}}, \bibinfo {author} {\bibfnamefont {G.~A.}\ \bibnamefont {Bocan}},
  \bibinfo {author} {\bibfnamefont {R.}~\bibnamefont {{D{\'\i}ez Mui{\~n}o}}},
  \ and\ \bibinfo {author} {\bibfnamefont {J.~I.}\ \bibnamefont {Juaristi}},\
  }\href {\doibase 10.1103/PhysRevLett.108.096101} {\bibfield  {journal}
  {\bibinfo  {journal} {Phys. Rev. Lett.}\ }\textbf {\bibinfo {volume} {108}},\
  \bibinfo {pages} {096101} (\bibinfo {year} {2012})}\BibitemShut {NoStop}%
\bibitem [{\citenamefont {F\"uchsel}\ \emph {et~al.}(2013)\citenamefont
  {F\"uchsel}, \citenamefont {Schimka},\ and\ \citenamefont
  {Saalfrank}}]{Fuechsel_JPhysChemA_2013}%
  \BibitemOpen
  \bibfield  {author} {\bibinfo {author} {\bibfnamefont {G.}~\bibnamefont
  {F\"uchsel}}, \bibinfo {author} {\bibfnamefont {S.}~\bibnamefont {Schimka}},
  \ and\ \bibinfo {author} {\bibfnamefont {P.}~\bibnamefont {Saalfrank}},\
  }\href {\doibase 10.1021/jp403860p} {\bibfield  {journal} {\bibinfo
  {journal} {J. Phys. Chem. A}\ }\textbf {\bibinfo {volume} {117}},\ \bibinfo
  {pages} {8761} (\bibinfo {year} {2013})}\BibitemShut {NoStop}%
\bibitem [{\citenamefont {Blanco-Rey}\ \emph {et~al.}(2014)\citenamefont
  {Blanco-Rey}, \citenamefont {Juaristi}, \citenamefont {{D\'{i}ez Mui\~no}},
  \citenamefont {Busnengo}, \citenamefont {Kroes},\ and\ \citenamefont
  {Alducin}}]{BlancoRey_PhysRevLett_2014}%
  \BibitemOpen
  \bibfield  {author} {\bibinfo {author} {\bibfnamefont {M.}~\bibnamefont
  {Blanco-Rey}}, \bibinfo {author} {\bibfnamefont {J.~I.}\ \bibnamefont
  {Juaristi}}, \bibinfo {author} {\bibfnamefont {R.}~\bibnamefont {{D\'{i}ez
  Mui\~no}}}, \bibinfo {author} {\bibfnamefont {H.~F.}\ \bibnamefont
  {Busnengo}}, \bibinfo {author} {\bibfnamefont {G.~J.}\ \bibnamefont {Kroes}},
  \ and\ \bibinfo {author} {\bibfnamefont {M.}~\bibnamefont {Alducin}},\ }\href
  {\doibase 10.1103/PhysRevLett.112.103203} {\bibfield  {journal} {\bibinfo
  {journal} {Phys. Rev. Lett.}\ }\textbf {\bibinfo {volume} {112}},\ \bibinfo
  {pages} {103203} (\bibinfo {year} {2014})}\BibitemShut {NoStop}%
\bibitem [{\citenamefont {Saalfrank}\ \emph {et~al.}(2014)\citenamefont
  {Saalfrank}, \citenamefont {Juaristi}, \citenamefont {Alducin}, \citenamefont
  {Blanco-Rey},\ and\ \citenamefont {{D{\'i}ez
  Mui{\~n}o}}}]{Saalfrank_JChemPhys_2014}%
  \BibitemOpen
  \bibfield  {author} {\bibinfo {author} {\bibfnamefont {P.}~\bibnamefont
  {Saalfrank}}, \bibinfo {author} {\bibfnamefont {J.~I.}\ \bibnamefont
  {Juaristi}}, \bibinfo {author} {\bibfnamefont {M.}~\bibnamefont {Alducin}},
  \bibinfo {author} {\bibfnamefont {M.}~\bibnamefont {Blanco-Rey}}, \ and\
  \bibinfo {author} {\bibfnamefont {R.}~\bibnamefont {{D{\'i}ez Mui{\~n}o}}},\
  }\href {\doibase http://dx.doi.org/10.1063/1.4903309} {\bibfield  {journal}
  {\bibinfo  {journal} {J. Chem. Phys.}\ }\textbf {\bibinfo {volume} {141}},\
  \bibinfo {eid} {234702} (\bibinfo {year} {2014})}\BibitemShut {NoStop}%
\bibitem [{\citenamefont {Li}\ and\ \citenamefont
  {Wahnstr\"om}(1992)}]{Li_PhysRevLett_1992}%
  \BibitemOpen
  \bibfield  {author} {\bibinfo {author} {\bibfnamefont {Y.}~\bibnamefont
  {Li}}\ and\ \bibinfo {author} {\bibfnamefont {G.}~\bibnamefont
  {Wahnstr\"om}},\ }\href {\doibase 10.1103/PhysRevLett.68.3444} {\bibfield
  {journal} {\bibinfo  {journal} {Phys. Rev. Lett.}\ }\textbf {\bibinfo
  {volume} {68}},\ \bibinfo {pages} {3444} (\bibinfo {year}
  {1992})}\BibitemShut {NoStop}%
\bibitem [{\citenamefont {Luntz}\ \emph {et~al.}(2009)\citenamefont {Luntz},
  \citenamefont {Makkonen}, \citenamefont {Persson}, \citenamefont {Holloway},
  \citenamefont {Bird},\ and\ \citenamefont
  {Mizielinski}}]{Luntz_PhysRevLett_2009}%
  \BibitemOpen
  \bibfield  {author} {\bibinfo {author} {\bibfnamefont {A.~C.}\ \bibnamefont
  {Luntz}}, \bibinfo {author} {\bibfnamefont {I.}~\bibnamefont {Makkonen}},
  \bibinfo {author} {\bibfnamefont {M.}~\bibnamefont {Persson}}, \bibinfo
  {author} {\bibfnamefont {S.}~\bibnamefont {Holloway}}, \bibinfo {author}
  {\bibfnamefont {D.~M.}\ \bibnamefont {Bird}}, \ and\ \bibinfo {author}
  {\bibfnamefont {M.~S.}\ \bibnamefont {Mizielinski}},\ }\href {\doibase
  10.1103/PhysRevLett.102.109601} {\bibfield  {journal} {\bibinfo  {journal}
  {Phys. Rev. Lett.}\ }\textbf {\bibinfo {volume} {102}},\ \bibinfo {pages}
  {109601} (\bibinfo {year} {2009})}\BibitemShut {NoStop}%
\bibitem [{\citenamefont {Juaristi}\ \emph {et~al.}(2009)\citenamefont
  {Juaristi}, \citenamefont {Alducin}, \citenamefont {{D\'{i}ez Mui\~no}},
  \citenamefont {Busnengo},\ and\ \citenamefont
  {Salin}}]{Juaristi_PhysRevLett_2009}%
  \BibitemOpen
  \bibfield  {author} {\bibinfo {author} {\bibfnamefont {J.~I.}\ \bibnamefont
  {Juaristi}}, \bibinfo {author} {\bibfnamefont {M.}~\bibnamefont {Alducin}},
  \bibinfo {author} {\bibfnamefont {R.}~\bibnamefont {{D\'{i}ez Mui\~no}}},
  \bibinfo {author} {\bibfnamefont {H.~F.}\ \bibnamefont {Busnengo}}, \ and\
  \bibinfo {author} {\bibfnamefont {A.}~\bibnamefont {Salin}},\ }\href
  {\doibase 10.1103/PhysRevLett.102.109602} {\bibfield  {journal} {\bibinfo
  {journal} {Phys. Rev. Lett.}\ }\textbf {\bibinfo {volume} {102}},\ \bibinfo
  {pages} {109602} (\bibinfo {year} {2009})}\BibitemShut {NoStop}%
\bibitem [{\citenamefont {Tully}\ \emph {et~al.}(1993)\citenamefont {Tully},
  \citenamefont {Gomez},\ and\ \citenamefont
  {Head-Gordon}}]{Tully_JVacSciTechnolA_1993}%
  \BibitemOpen
  \bibfield  {author} {\bibinfo {author} {\bibfnamefont {J.~C.}\ \bibnamefont
  {Tully}}, \bibinfo {author} {\bibfnamefont {M.}~\bibnamefont {Gomez}}, \ and\
  \bibinfo {author} {\bibfnamefont {M.}~\bibnamefont {Head-Gordon}},\ }\href
  {\doibase 10.1116/1.578522} {\bibfield  {journal} {\bibinfo  {journal} {J.
  Vac. Sci. Technol. A}\ }\textbf {\bibinfo {volume} {11}},\ \bibinfo {pages}
  {1914} (\bibinfo {year} {1993})}\BibitemShut {NoStop}%
\bibitem [{\citenamefont {Arnolds}(2011)}]{Arnolds_ProgSurfSci_2011}%
  \BibitemOpen
  \bibfield  {author} {\bibinfo {author} {\bibfnamefont {H.}~\bibnamefont
  {Arnolds}},\ }\href {\doibase 10.1016/j.progsurf.2010.10.001} {\bibfield
  {journal} {\bibinfo  {journal} {Prog. Surf. Sci.}\ }\textbf {\bibinfo
  {volume} {86}},\ \bibinfo {pages} {1 } (\bibinfo {year} {2011})}\BibitemShut
  {NoStop}%
\bibitem [{\citenamefont {Saalfrank}(2006)}]{Saalfrank_ChemRev_2006}%
  \BibitemOpen
  \bibfield  {author} {\bibinfo {author} {\bibfnamefont {P.}~\bibnamefont
  {Saalfrank}},\ }\href {\doibase 10.1021/cr0501691} {\bibfield  {journal}
  {\bibinfo  {journal} {Chem. Rev.}\ }\textbf {\bibinfo {volume} {106}},\
  \bibinfo {pages} {4116} (\bibinfo {year} {2006})}\BibitemShut {NoStop}%
\bibitem [{\citenamefont {Beckerle}\ \emph {et~al.}(1991)\citenamefont
  {Beckerle}, \citenamefont {Cavanagh}, \citenamefont {Casassa}, \citenamefont
  {Heilweil},\ and\ \citenamefont {Stephenson}}]{Beckerle_JChemPhys_1991}%
  \BibitemOpen
  \bibfield  {author} {\bibinfo {author} {\bibfnamefont {J.~D.}\ \bibnamefont
  {Beckerle}}, \bibinfo {author} {\bibfnamefont {R.~R.}\ \bibnamefont
  {Cavanagh}}, \bibinfo {author} {\bibfnamefont {M.~P.}\ \bibnamefont
  {Casassa}}, \bibinfo {author} {\bibfnamefont {E.~J.}\ \bibnamefont
  {Heilweil}}, \ and\ \bibinfo {author} {\bibfnamefont {J.~C.}\ \bibnamefont
  {Stephenson}},\ }\href {\doibase 10.1063/1.461657} {\bibfield  {journal}
  {\bibinfo  {journal} {J. Chem. Phys.}\ }\textbf {\bibinfo {volume} {95}},\
  \bibinfo {pages} {5403} (\bibinfo {year} {1991})}\BibitemShut {NoStop}%
\bibitem [{\citenamefont {Morin}\ \emph {et~al.}(1992)\citenamefont {Morin},
  \citenamefont {Levinos},\ and\ \citenamefont
  {Harris}}]{Morin_JChemPhys_1992}%
  \BibitemOpen
  \bibfield  {author} {\bibinfo {author} {\bibfnamefont {M.}~\bibnamefont
  {Morin}}, \bibinfo {author} {\bibfnamefont {N.~J.}\ \bibnamefont {Levinos}},
  \ and\ \bibinfo {author} {\bibfnamefont {A.~L.}\ \bibnamefont {Harris}},\
  }\href {\doibase 10.1063/1.461897} {\bibfield  {journal} {\bibinfo  {journal}
  {J. Chem. Phys.}\ }\textbf {\bibinfo {volume} {96}},\ \bibinfo {pages} {3950}
  (\bibinfo {year} {1992})}\BibitemShut {NoStop}%
\bibitem [{\citenamefont {Forsblom}\ and\ \citenamefont
  {Persson}(2007)}]{Forsblom_JChemPhys_2007}%
  \BibitemOpen
  \bibfield  {author} {\bibinfo {author} {\bibfnamefont {M.}~\bibnamefont
  {Forsblom}}\ and\ \bibinfo {author} {\bibfnamefont {M.}~\bibnamefont
  {Persson}},\ }\href {\doibase 10.1063/1.2794744} {\bibfield  {journal}
  {\bibinfo  {journal} {J. Chem. Phys.}\ }\textbf {\bibinfo {volume} {127}},\
  \bibinfo {eid} {154303} (\bibinfo {year} {2007})}\BibitemShut {NoStop}%
\bibitem [{\citenamefont {Krishna}\ and\ \citenamefont
  {Tully}(2006)}]{Krishna_JChemPhys_2006}%
  \BibitemOpen
  \bibfield  {author} {\bibinfo {author} {\bibfnamefont {V.}~\bibnamefont
  {Krishna}}\ and\ \bibinfo {author} {\bibfnamefont {J.~C.}\ \bibnamefont
  {Tully}},\ }\href {\doibase 10.1063/1.2227383} {\bibfield  {journal}
  {\bibinfo  {journal} {J. Chem. Phys.}\ }\textbf {\bibinfo {volume} {125}},\
  \bibinfo {eid} {054706} (\bibinfo {year} {2006})}\BibitemShut {NoStop}%
\bibitem [{\citenamefont {Persson}\ and\ \citenamefont
  {Hellsing}(1982)}]{Persson_PhysRevLett_1982}%
  \BibitemOpen
  \bibfield  {author} {\bibinfo {author} {\bibfnamefont {M.}~\bibnamefont
  {Persson}}\ and\ \bibinfo {author} {\bibfnamefont {B.}~\bibnamefont
  {Hellsing}},\ }\href {\doibase 10.1103/PhysRevLett.49.662} {\bibfield
  {journal} {\bibinfo  {journal} {Phys. Rev. Lett.}\ }\textbf {\bibinfo
  {volume} {49}},\ \bibinfo {pages} {662} (\bibinfo {year} {1982})}\BibitemShut
  {NoStop}%
\bibitem [{\citenamefont {Lorente}\ and\ \citenamefont
  {Persson}(2000)}]{Lorente_FaradayDiscuss_2000}%
  \BibitemOpen
  \bibfield  {author} {\bibinfo {author} {\bibfnamefont {N.}~\bibnamefont
  {Lorente}}\ and\ \bibinfo {author} {\bibfnamefont {M.}~\bibnamefont
  {Persson}},\ }\href {\doibase 10.1039/B002826F} {\bibfield  {journal}
  {\bibinfo  {journal} {Faraday Discuss.}\ }\textbf {\bibinfo {volume} {117}},\
  \bibinfo {pages} {277} (\bibinfo {year} {2000})}\BibitemShut {NoStop}%
\bibitem [{\citenamefont {Luntz}\ \emph {et~al.}(2006)\citenamefont {Luntz},
  \citenamefont {Persson}, \citenamefont {Wagner}, \citenamefont {Frischkorn},\
  and\ \citenamefont {Wolf}}]{Luntz_JChemPhys_2006}%
  \BibitemOpen
  \bibfield  {author} {\bibinfo {author} {\bibfnamefont {A.~C.}\ \bibnamefont
  {Luntz}}, \bibinfo {author} {\bibfnamefont {M.}~\bibnamefont {Persson}},
  \bibinfo {author} {\bibfnamefont {S.}~\bibnamefont {Wagner}}, \bibinfo
  {author} {\bibfnamefont {C.}~\bibnamefont {Frischkorn}}, \ and\ \bibinfo
  {author} {\bibfnamefont {M.}~\bibnamefont {Wolf}},\ }\href {\doibase
  http://dx.doi.org/10.1063/1.2206588} {\bibfield  {journal} {\bibinfo
  {journal} {J. Chem. Phys.}\ }\textbf {\bibinfo {volume} {124}},\ \bibinfo
  {eid} {244702} (\bibinfo {year} {2006})}\BibitemShut {NoStop}%
\bibitem [{\citenamefont {Puska}\ and\ \citenamefont
  {Nieminen}(1983)}]{Puska_PhysRevB_1983}%
  \BibitemOpen
  \bibfield  {author} {\bibinfo {author} {\bibfnamefont {M.~J.}\ \bibnamefont
  {Puska}}\ and\ \bibinfo {author} {\bibfnamefont {R.~M.}\ \bibnamefont
  {Nieminen}},\ }\href {\doibase 10.1103/PhysRevB.27.6121} {\bibfield
  {journal} {\bibinfo  {journal} {Phys. Rev. B}\ }\textbf {\bibinfo {volume}
  {27}},\ \bibinfo {pages} {6121} (\bibinfo {year} {1983})}\BibitemShut
  {NoStop}%
\bibitem [{\citenamefont {Clark}\ \emph {et~al.}(2005)\citenamefont {Clark},
  \citenamefont {Segall}, \citenamefont {Pickard}, \citenamefont {Hasnip},
  \citenamefont {Probert}, \citenamefont {Refson},\ and\ \citenamefont
  {Payne}}]{Clark_ZKristallogr_2005}%
  \BibitemOpen
  \bibfield  {author} {\bibinfo {author} {\bibfnamefont {S.~J.}\ \bibnamefont
  {Clark}}, \bibinfo {author} {\bibfnamefont {M.~D.}\ \bibnamefont {Segall}},
  \bibinfo {author} {\bibfnamefont {C.~J.}\ \bibnamefont {Pickard}}, \bibinfo
  {author} {\bibfnamefont {P.~J.}\ \bibnamefont {Hasnip}}, \bibinfo {author}
  {\bibfnamefont {M.~I.~J.}\ \bibnamefont {Probert}}, \bibinfo {author}
  {\bibfnamefont {K.}~\bibnamefont {Refson}}, \ and\ \bibinfo {author}
  {\bibfnamefont {M.~C.}\ \bibnamefont {Payne}},\ }\href {\doibase
  10.1524/zkri.220.5.567.65075} {\bibfield  {journal} {\bibinfo  {journal} {Z.
  Kristallogr.}\ }\textbf {\bibinfo {volume} {220}},\ \bibinfo {pages} {567}
  (\bibinfo {year} {2005})}\BibitemShut {NoStop}%
\bibitem [{\citenamefont {Perdew}\ \emph {et~al.}(1996)\citenamefont {Perdew},
  \citenamefont {Burke},\ and\ \citenamefont
  {Ernzerhof}}]{Perdew_PhysRevLett_1996}%
  \BibitemOpen
  \bibfield  {author} {\bibinfo {author} {\bibfnamefont {J.~P.}\ \bibnamefont
  {Perdew}}, \bibinfo {author} {\bibfnamefont {K.}~\bibnamefont {Burke}}, \
  and\ \bibinfo {author} {\bibfnamefont {M.}~\bibnamefont {Ernzerhof}},\ }\href
  {\doibase 10.1103/PhysRevLett.77.3865} {\bibfield  {journal} {\bibinfo
  {journal} {Phys. Rev. Lett.}\ }\textbf {\bibinfo {volume} {77}},\ \bibinfo
  {pages} {3865} (\bibinfo {year} {1996})}\BibitemShut {NoStop}%
\bibitem [{\citenamefont {Perdew}\ \emph {et~al.}(1997)\citenamefont {Perdew},
  \citenamefont {Burke},\ and\ \citenamefont
  {Ernzerhof}}]{Perdew_PhysRevLett_1997}%
  \BibitemOpen
  \bibfield  {author} {\bibinfo {author} {\bibfnamefont {J.~P.}\ \bibnamefont
  {Perdew}}, \bibinfo {author} {\bibfnamefont {K.}~\bibnamefont {Burke}}, \
  and\ \bibinfo {author} {\bibfnamefont {M.}~\bibnamefont {Ernzerhof}},\ }\href
  {\doibase 10.1103/PhysRevLett.78.1396} {\bibfield  {journal} {\bibinfo
  {journal} {Phys. Rev. Lett.}\ }\textbf {\bibinfo {volume} {78}},\ \bibinfo
  {pages} {1396} (\bibinfo {year} {1997})}\BibitemShut {NoStop}%
\bibitem [{\citenamefont {Vanderbilt}(1990)}]{Vanderbilt_PhysRevB_1990}%
  \BibitemOpen
  \bibfield  {author} {\bibinfo {author} {\bibfnamefont {D.}~\bibnamefont
  {Vanderbilt}},\ }\href {\doibase 10.1103/PhysRevB.41.7892} {\bibfield
  {journal} {\bibinfo  {journal} {Phys. Rev. B}\ }\textbf {\bibinfo {volume}
  {41}},\ \bibinfo {pages} {7892} (\bibinfo {year} {1990})}\BibitemShut
  {NoStop}%
\bibitem [{\citenamefont {Monkhorst}\ and\ \citenamefont
  {Pack}(1976)}]{Monkhorst_PhysRevB_1976}%
  \BibitemOpen
  \bibfield  {author} {\bibinfo {author} {\bibfnamefont {H.~J.}\ \bibnamefont
  {Monkhorst}}\ and\ \bibinfo {author} {\bibfnamefont {J.~D.}\ \bibnamefont
  {Pack}},\ }\href {\doibase 10.1103/PhysRevB.13.5188} {\bibfield  {journal}
  {\bibinfo  {journal} {Phys. Rev. B}\ }\textbf {\bibinfo {volume} {13}},\
  \bibinfo {pages} {5188} (\bibinfo {year} {1976})}\BibitemShut {NoStop}%
\bibitem [{\citenamefont {Feibelman}\ \emph {et~al.}(2001)\citenamefont
  {Feibelman}, \citenamefont {Hammer}, \citenamefont {N\o{}rskov},
  \citenamefont {Wagner}, \citenamefont {Scheffler}, \citenamefont {Stumpf},
  \citenamefont {Watwe},\ and\ \citenamefont
  {Dumesic}}]{Feibelman_JPhysChem_2001}%
  \BibitemOpen
  \bibfield  {author} {\bibinfo {author} {\bibfnamefont {P.~J.}\ \bibnamefont
  {Feibelman}}, \bibinfo {author} {\bibfnamefont {B.}~\bibnamefont {Hammer}},
  \bibinfo {author} {\bibfnamefont {J.~K.}\ \bibnamefont {N\o{}rskov}},
  \bibinfo {author} {\bibfnamefont {F.}~\bibnamefont {Wagner}}, \bibinfo
  {author} {\bibfnamefont {M.}~\bibnamefont {Scheffler}}, \bibinfo {author}
  {\bibfnamefont {R.}~\bibnamefont {Stumpf}}, \bibinfo {author} {\bibfnamefont
  {R.}~\bibnamefont {Watwe}}, \ and\ \bibinfo {author} {\bibfnamefont
  {J.}~\bibnamefont {Dumesic}},\ }\href {\doibase 10.1021/jp002302t} {\bibfield
   {journal} {\bibinfo  {journal} {J. Phys. Chem. B}\ }\textbf {\bibinfo
  {volume} {105}},\ \bibinfo {pages} {4018} (\bibinfo {year}
  {2001})}\BibitemShut {NoStop}%
\bibitem [{\citenamefont {Dion}\ \emph {et~al.}(2004)\citenamefont {Dion},
  \citenamefont {Rydberg}, \citenamefont {Schr\"oder}, \citenamefont
  {Langreth},\ and\ \citenamefont {Lundqvist}}]{Dion_PhysRevLett_2004}%
  \BibitemOpen
  \bibfield  {author} {\bibinfo {author} {\bibfnamefont {M.}~\bibnamefont
  {Dion}}, \bibinfo {author} {\bibfnamefont {H.}~\bibnamefont {Rydberg}},
  \bibinfo {author} {\bibfnamefont {E.}~\bibnamefont {Schr\"oder}}, \bibinfo
  {author} {\bibfnamefont {D.~C.}\ \bibnamefont {Langreth}}, \ and\ \bibinfo
  {author} {\bibfnamefont {B.~I.}\ \bibnamefont {Lundqvist}},\ }\href {\doibase
  10.1103/PhysRevLett.92.246401} {\bibfield  {journal} {\bibinfo  {journal}
  {Phys. Rev. Lett.}\ }\textbf {\bibinfo {volume} {92}},\ \bibinfo {pages}
  {246401} (\bibinfo {year} {2004})}\BibitemShut {NoStop}%
\bibitem [{\citenamefont {Lazi\'{c}}\ \emph {et~al.}(2010)\citenamefont
  {Lazi\'{c}}, \citenamefont {Alaei}, \citenamefont {Atodiresei}, \citenamefont
  {Caciuc}, \citenamefont {Brako},\ and\ \citenamefont
  {Bl\"ugel}}]{Lazic_PhysRevB_2010}%
  \BibitemOpen
  \bibfield  {author} {\bibinfo {author} {\bibfnamefont {P.}~\bibnamefont
  {Lazi\'{c}}}, \bibinfo {author} {\bibfnamefont {M.}~\bibnamefont {Alaei}},
  \bibinfo {author} {\bibfnamefont {N.}~\bibnamefont {Atodiresei}}, \bibinfo
  {author} {\bibfnamefont {V.}~\bibnamefont {Caciuc}}, \bibinfo {author}
  {\bibfnamefont {R.}~\bibnamefont {Brako}}, \ and\ \bibinfo {author}
  {\bibfnamefont {S.}~\bibnamefont {Bl\"ugel}},\ }\href {\doibase
  10.1103/PhysRevB.81.045401} {\bibfield  {journal} {\bibinfo  {journal} {Phys.
  Rev. B}\ }\textbf {\bibinfo {volume} {81}},\ \bibinfo {pages} {045401}
  (\bibinfo {year} {2010})}\BibitemShut {NoStop}%
\bibitem [{\citenamefont {Winter}\ \emph {et~al.}(2003)\citenamefont {Winter},
  \citenamefont {Juaristi}, \citenamefont {Nagy}, \citenamefont {Arnau},\ and\
  \citenamefont {Echenique}}]{Winter_PhysRevB_2003}%
  \BibitemOpen
  \bibfield  {author} {\bibinfo {author} {\bibfnamefont {H.}~\bibnamefont
  {Winter}}, \bibinfo {author} {\bibfnamefont {J.~I.}\ \bibnamefont
  {Juaristi}}, \bibinfo {author} {\bibfnamefont {I.}~\bibnamefont {Nagy}},
  \bibinfo {author} {\bibfnamefont {A.}~\bibnamefont {Arnau}}, \ and\ \bibinfo
  {author} {\bibfnamefont {P.~M.}\ \bibnamefont {Echenique}},\ }\href {\doibase
  10.1103/PhysRevB.67.245401} {\bibfield  {journal} {\bibinfo  {journal} {Phys.
  Rev. B}\ }\textbf {\bibinfo {volume} {67}},\ \bibinfo {pages} {245401}
  (\bibinfo {year} {2003})}\BibitemShut {NoStop}%
\bibitem [{\citenamefont {Tremblay}\ \emph {et~al.}(2010)\citenamefont
  {Tremblay}, \citenamefont {Monturet},\ and\ \citenamefont
  {Saalfrank}}]{Tremblay_PhysRevB_2010}%
  \BibitemOpen
  \bibfield  {author} {\bibinfo {author} {\bibfnamefont {J.~C.}\ \bibnamefont
  {Tremblay}}, \bibinfo {author} {\bibfnamefont {S.}~\bibnamefont {Monturet}},
  \ and\ \bibinfo {author} {\bibfnamefont {P.}~\bibnamefont {Saalfrank}},\
  }\href {\doibase 10.1103/PhysRevB.81.125408} {\bibfield  {journal} {\bibinfo
  {journal} {Phys. Rev. B}\ }\textbf {\bibinfo {volume} {81}},\ \bibinfo
  {pages} {125408} (\bibinfo {year} {2010})}\BibitemShut {NoStop}%
\bibitem [{\citenamefont {Hirshfeld}(1977)}]{Hirshfeld_TheoretChimActa_1977}%
  \BibitemOpen
  \bibfield  {author} {\bibinfo {author} {\bibfnamefont {F.}~\bibnamefont
  {Hirshfeld}},\ }\href {\doibase 10.1007/BF00549096} {\bibfield  {journal}
  {\bibinfo  {journal} {Theoret. Chim. Acta}\ }\textbf {\bibinfo {volume}
  {44}},\ \bibinfo {pages} {129} (\bibinfo {year} {1977})}\BibitemShut
  {NoStop}%
\bibitem [{\citenamefont {McNellis}\ \emph {et~al.}(2009)\citenamefont
  {McNellis}, \citenamefont {Meyer},\ and\ \citenamefont
  {Reuter}}]{McNellis_PhysRevB_2009}%
  \BibitemOpen
  \bibfield  {author} {\bibinfo {author} {\bibfnamefont {E.~R.}\ \bibnamefont
  {McNellis}}, \bibinfo {author} {\bibfnamefont {J.}~\bibnamefont {Meyer}}, \
  and\ \bibinfo {author} {\bibfnamefont {K.}~\bibnamefont {Reuter}},\ }\href
  {\doibase 10.1103/PhysRevB.80.205414} {\bibfield  {journal} {\bibinfo
  {journal} {Phys. Rev. B}\ }\textbf {\bibinfo {volume} {80}},\ \bibinfo
  {pages} {205414} (\bibinfo {year} {2009})}\BibitemShut {NoStop}%
\bibitem [{Note1()}]{Note1}%
  \BibitemOpen
  \bibinfo {note} {Consistent with the theoretical work by FP \cite
  {Forsblom_JChemPhys_2007}, we consider perpendicular CN adsorption with the C
  atom coordinated to the metal on a top site within a $(2\times 2)$ surface
  unit cell. The H$_2$ on Ru(0001) system is modeled by analogy with the
  equilibrium configuration mentioned by F{\"u}chsel \protect \emph {et al.}.
  \cite {Fuechsel_JPhysChemA_2013}, i.e., using a $(2\times 2)$ surface unit
  cell with two H atoms adsorbed in adjacent fcc hollow sites. For both
  systems, we use a $(9\times 9\times 1)$ Monkhorst-Pack $\protect \mathbf
  {k}$-point grid \cite {Monkhorst_PhysRevB_1976}. All remaining computational
  details are the same as outlined for CO on Cu(100) and Pt(111),
  respectively.}\BibitemShut {Stop}%
\bibitem [{\citenamefont {Meyer}\ and\ \citenamefont
  {Reuter}(2014)}]{Meyer_AngewChemIntEd_2014}%
  \BibitemOpen
  \bibfield  {author} {\bibinfo {author} {\bibfnamefont {J.}~\bibnamefont
  {Meyer}}\ and\ \bibinfo {author} {\bibfnamefont {K.}~\bibnamefont {Reuter}},\
  }\href {\doibase 10.1002/anie.201400066} {\bibfield  {journal} {\bibinfo
  {journal} {Angew. Chem. Int. Ed.}\ }\textbf {\bibinfo {volume} {53}},\
  \bibinfo {pages} {4721} (\bibinfo {year} {2014})}\BibitemShut {NoStop}%
\end{thebibliography}%
\end{document}